# Tracking the Ultrafast Non-Equilibrium Energy Flow between Electronic and Lattice Degrees of Freedom in Crystalline Nickel

P. Maldonado[1], T. Chase[2,3], A. H. Reid[2], X. Shen[2], R. K. Li[2], K. Carva[4], T. Payer[5], M. Horn von Hoegen[5], K. Sokolowski-Tinten[5], X.J. Wang[2], P.M. Oppeneer[1], and H.A. Dürr[1,2]

[1] *Department of Physics and Astronomy, Uppsala University, P. O. Box 516, S-75120 Uppsala, Sweden*
[2] *SLAC National Accelerator Laboratory, 2575 Sand Hill Road, Menlo Park, CA 94025, USA*
[3] *Department of Applied Physics, Stanford University, Stanford, California 94305, USA*
[4] *Faculty of Mathematics and Physics, Department of Condensed Matter Physics, Charles University, Ke Karlovu 5, CZ-12116 Prague 2, Czech Republic*
[5] *Department of Physics and Center for Nanointegration Duisburg-Essen (CENIDE), University of Duisburg-Essen, 47057 Duisburg, Germany*

Femtosecond laser excitation of solid-state systems creates non-equilibrium hot electrons that cool down by transferring their energy to other degrees of freedom and ultimately to lattice vibrations of the solid. By combining *ab initio* calculations with ultrafast diffuse electron scattering we gain a detailed understanding of the complex non-equilibrium energy transfer between electrons and phonons in laser-excited Ni metal. Our experimental results show that the wavevector resolved population dynamics of phonon modes is distinctly different throughout the Brillouin zone and are in remarkable agreement with our theoretical results. We find that zone-boundary phonon modes become occupied first. As soon as the energy in these modes becomes larger than the average electron energy a backflow of energy from lattice to electronic degrees of freedom occurs. Subsequent excitation of lower-energy phonon modes drives the thermalization of the whole system on the picosecond timescale. We determine the evolving non-equilibrium phonon occupations which we find to deviate markedly from thermal occupations.

*Introduction.* The ability to measure dynamical processes in real time with femtosecond time-resolution has in recent years enabled observations of unexpected, non-equilibrium dynamical phenomena [1-4]. In such pump-probe measurements an ultrashort optical laser pulse brings the electron system impulsively into a highly non-equilibrium state which is followed by energy transfer to other (lattice, spin) degrees of freedom in the solid. [5-7]. The ensuing strongly non-equilibrium dynamics offers innovative non-thermo-dynamic pathways to achieve ultrafast control of the material's properties already in the initial state before thermalization has been reached on longer timescales [8-12].

The understanding of strongly non-equilibrium dynamics in solids is still very limited, in spite of its emerging importance from a fundamental and applied science viewpoint. The two-temperature model (2TM) [13], commonly used for metals, assumes that the electronic and phononic subsystems are each in separate equilibrium at all times and reach global equilibrium by exchanging heat [13, 14]. Recent experiments have revealed that these assumptions are no longer tenable on ultrashort timescales [15-19]. The excited electron system reaches rapidly a uniform electron temperature through fast electron-electron scattering within a few hundred femtoseconds [20,21]. However, even in "simple" systems (e.g. mono-atomic metals) the equilibration of the phononic system proceeds along a complex pathway and it can take tens of picoseconds before a homogeneous temperature is reached [7,17,22].

Time-resolved diffraction techniques are commonly used to determine atomic vibration amplitudes averaged over the Brillouin Zone (BZ) via Debye-Waller analysis of Bragg peaks [7,16,17,23-26]. However, in order to disentangle the highly correlated non-equilibrium dynamics of electrons and phonons throughout the whole BZ, precise measurements of the non-thermal momentum-dependent phonon populations on the (sub)picosecond timescale are required. Such information could be obtained from *diffuse* ultrafast X-ray and electron diffraction [7,18,27-29]. The full potential of this technique has however not yet been explored, because it requires additional detailed knowledge of the evolving energy flow. This knowledge can be obtained using a recently proposed materials' specific *ab initio* theory of the non-equilibrium dynamics [22]. The combination of experiment and theory is thus poised to unlock new possibilities to track energy flow between excited electrons and phonons offering a detailed picture of the non-equilibrium phonon dynamics.

In this letter, we investigate the ultrafast non-equilibrium dynamics in a single-crystalline Ni. We use diffuse ultrafast electron diffraction (UED) [7,18, 29] to track the momentum-resolved phonon occupation dynamics throughout the BZ, which we find to be distinctly different from a thermalized occupation for the first 5 ps after laser excitation. Non-equilibrium *ab initio* theory [22] provides a parameter free description of the phonon dynamics in good agreement with our measurements. This enables us to disentangle the electron-phonon (el-ph) and phonon-phonon (ph-ph) energy transfer processes. We identify a previously unnoted thermalization process, the backflow of energy from zone-boundary phonon modes to the electrons. In addition, we find that changes in the electronic structure due to ultrafast demagnetization [20] are needed to quantitatively model the observed electron-phonon energy transfer.

*Experiment.* UED in a pump-probe setup [7,30] is used to investigate the temporal evolution of the non-equilibrium phonon populations of a single-crystalline Ni film upon femtosecond laser excitation. The reduced multiple scattering at 3.3 MeV beam energy and a flat Ewald-sphere are critical to compare our experiments to *ab initio* calculations. 20 nm thick Ni (001) films were grown by molecular beam epitaxy onto single-crystalline NaCl (001). The substrate was subsequently dissolved and the Ni films were floated onto TEM grids. This process results in free-standing single-crystalline Ni films that display sharp Bragg peaks (see Fig. 1(b)). Near-infrared pump pulses at 800 nm central wavelength and 40 fs duration bring the Ni valence electrons into a highly non-equilibrium state. The size of the pump laser spot was adjusted to provide homogeneous illumination of the Ni sample over the electron beam size of ~100 µm. The laser-excited Ni valence electrons will transfer energy to the phonons through electron-phonon interaction as shown schematically in Fig. 1(a). The response of the phonons is measured in real-time with UED at the SLAC UED facility [30] (200 fs pulse duration) in transmission as shown schematically in the inset of Fig. 1(b). Time-resolved scattering patterns are recorded for several pump-probe time delays up to 5 ps.

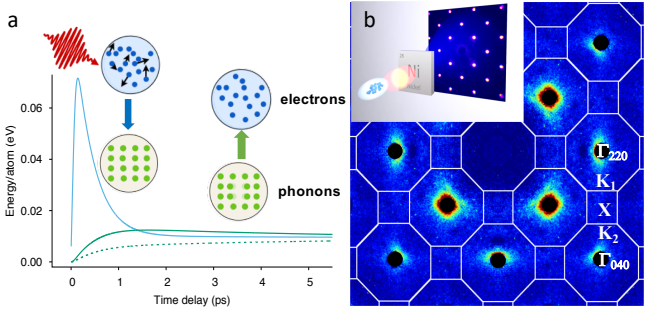

**Fig. 1**. (a) Sketch of the energy transfer between electronic system and lattice vibrations. Shown are the calculated atomic electron energy (blue line) and vibrational energy for phonon modes close to the X (solid green line) and Γ (dashed green line) points. (b) Measured diffuse scattering intensity for the experimental geometry that is schematically shown in the inset. The figure shows the difference intensity between pump-laser on and off. Diffuse intensities have been averaged over equivalent regions of the four-fold symmetric diffraction pattern. The regions of the Bragg peaks are set to zero (black discs). Brillouin zone boundaries and selected high-symmetry points are marked in white.

Figure 1(b) shows an example of a scattering pattern measured at 1 ps pump-probe time delay. The diffuse intensity is many orders of magnitude lower than that of the Bragg peaks and is therefore only visible if the stronger Bragg scattering is removed (black discs in Fig. 1(b)). White lines depict the boundaries of the BZ of fcc Ni and letters specify the positions of high-symmetry points in reciprocal space.

The time evolution of the diffuse scattering intensity [7] provides information about the non-equilibrium phonon populations, in a momentum resolved way. The diffuse intensity in reciprocal space is given as [31,32],

$$I(\mathbf{Q}) \propto \sum_\nu \frac{1}{\omega_\nu(\mathbf{q})}\left[n_\nu(\mathbf{q}) + \frac{1}{2}\right]|F_\nu(\mathbf{Q})|^2, \qquad (1)$$

where $\omega_\nu(\mathbf{q})$ is the frequency of the phonon with mode $\nu$ and reduced wavevector $\mathbf{q}$, $n_\nu(\mathbf{q})$ is the phonon occupation number and $F$ is the structure factor, $F_\nu(\mathbf{Q}) \propto \sum_s \frac{f_s}{\sqrt{m}} e^{-M_s}[\mathbf{Q} \cdot \boldsymbol{\varepsilon}_\nu] e^{-i(\mathbf{K}_Q \cdot \mathbf{r}_s)}$ [31,32]. In this expression, $f$ is the atomic scattering factor, $m$ and $M_s$ are the mass and the Debye-Waller factor, respectively, of atom $s$ at position $\mathbf{r}_s$, $\boldsymbol{\varepsilon}_\nu$ is the phonon polarization vector, and $\mathbf{K}_Q$ is the closest reciprocal-lattice vector to $\mathbf{Q}$, i.e., $\mathbf{Q} = \mathbf{q} + \mathbf{K}_Q$. In the following we focus on the diffuse scattering evolution along the $\Gamma_{220}$-X-$\Gamma_{400}$ high-symmetry line spanned by the $\mathbf{K}_{220}$ and $\mathbf{K}_{400}$ reciprocal lattice vectors (see Fig. 1(b)). We have therefore averaged over all four equivalent quadrants of the diffraction pattern shown in Fig. 1(b). Equation (1) shows that along $\Gamma_{220}$-X-$\Gamma_{400}$ the projection, $\mathbf{Q} \cdot \boldsymbol{\varepsilon}_\nu$, of the phonon polarization, $\boldsymbol{\varepsilon}_\nu$, onto the total transferred wavevector, $\mathbf{Q}$, changes. This implies that close to $\Gamma_{220}$ we mainly probe phonons with a polarization transverse to their wavevectors $\mathbf{q}$, while near $\Gamma_{400}$ we are equally sensitive to transverse and longitudinally polarized phonons. In the chosen scattering geometry, we are not sensitive to phonons with transverse polarization perpendicular to the plane shown in Fig. 1(b).

*Theory.* Our recently developed non-equilibrium theory of the wavevector-dependent ultrafast electron and lattice dynamics [22] has been extended here to explicitly provide the energy transfer between the laser-excited electrons and the lattice. The main features are the phonon branch and wavevector dependence of e-ph coupling and an explicit inclusion of anharmonic effects describing ph-ph scattering events. We model the non-equilibrium variation of the wavevector-dependent phonon populations as seen by UED [7]. For the employed laser excitation fluences far below the melting threshold, any variation of the diffuse scattering intensity due to phonon mode softening (see Eq. (1)) is expected to be small [33] and is hence neglected. Importantly, our kinetic theory captures the full transient dynamics of the non-equilibrium phononic populations. The rate of exchange that defines the time evolution of the non-equilibrium energy flow between the electronic system and the different phonon modes after laser excitation is calculated by numerically solving the following rate equations

$$\frac{\partial E_e}{\partial t} = \sum_{\mathbf{q},\nu} \hbar\omega_\nu(\mathbf{q})\gamma_\nu(\mathbf{q}, E_e, t)[n_\nu(\mathbf{q}, E_l^\mathbf{q}) - n_\nu(\mathbf{q}, E_e)] + P(t),$$

$$\frac{\partial E_\nu^\mathbf{q}}{\partial t} = -\hbar\omega_\nu(\mathbf{q})\gamma_\nu(\mathbf{q}, E_e, t)[n_\nu(\mathbf{q}, E_l^\mathbf{q}) - n_\nu(\mathbf{q}, E_e)] + \frac{\partial E^{ph-ph}}{\partial t}$$
$$\text{for } \mathbf{q} = \mathbf{q}_1, \cdots, \mathbf{q}_N \qquad (2)$$

where $n_\nu(\mathbf{q}, E_l^\mathbf{q})$ is the non-equilibrium phonon population of phonon mode $\mathbf{q}$ with branch $\nu$, with $E_l^\mathbf{q}$ being the time-dependent amount of energy stored in this particular mode. $P(t)$ is the pump laser field that generates the non-equilibrium electronic distribution. $\gamma_\nu(\mathbf{q}, E_e, t)$ is the phonon linewidth due to e-ph scattering, which depends explicitly on the phonon mode, electronic spin degrees of freedom and on the electronic energy, $E_e$. It is therefore a time dependent quantity. Note that, while the first terms on the right-hand side of Eqs. (2) define the energy flow due to e-ph interaction, the term, $\frac{\partial E^{ph-ph}}{\partial t}$, defines the energy flow due to ph-ph scattering processes and explicitly accounts for the system anharmonicities [22]. To obtain a full solution of the non-equilibrium model defined by Eq. (2), we compute all required material-specific quantities using spin-polarized density functional theory, and solve Eq. (2) numerically [22].

*Results.* Figure 2 shows measured and calculated Ni phonon properties for phonon wavevectors along the $\Gamma_{220}$-X-$\Gamma_{400}$ high-symmetry line, between the (220) and (400) Bragg peaks in Fig. 1(b). Figure 2(a) displays the calculated phonon dispersions along this line. Experimentally only the L and T phonon modes in Fig 2(a) with a polarization in the scattering plane shown in Fig. 1(b) can be detected. The $T_{perp}$ mode (dotted line in Fig. 2(a)) is polarized perpendicular to this plane and is therefore not observed. The measured diffuse scattering intensities for different time delays are displayed in Fig. 2(b) (symbols) with vertical offsets for clarity. The curves represent the difference intensity between laser-on and laser-off at the indicated pump-probe time delays and therefore correspond to the laser-induced changes in the phonon population (see Eq. (1)). Note that on approaching the Γ–points the measured intensities are dominated by Debye-Waller attenuation of the Bragg peaks [7]. These data are omitted in Fig. 2(b). The diffuse intensities display characteristic changes with time delay: The total amount of scattering intensity increases and the intensity distribution shifts towards the Γ-points. This is qualitatively similar to our previous observations in Au [7] but the changes occur much faster for Ni.

Lines in Fig. 2(b) represent two calculated scenarios, transient phonon excitation for ferromagnetic Ni (dashed lines) and for a ferromagnetic to non-magnetic phase transition (drawn lines). It is evident that the purely ferromagnetic case does not reproduce the experimental observations. In this case there is a fast transfer of electronic energy into phonons resulting in a much higher mode occupation, and also a faster lattice relaxation, than experimentally observed at all pump-probe delay times. Since we are using *ab initio* electron-phonon coupling parameters we can conclude that assuming an unchanged e-ph coupling of Ni obviously does not properly describe the case of ultrafast laser excitation.

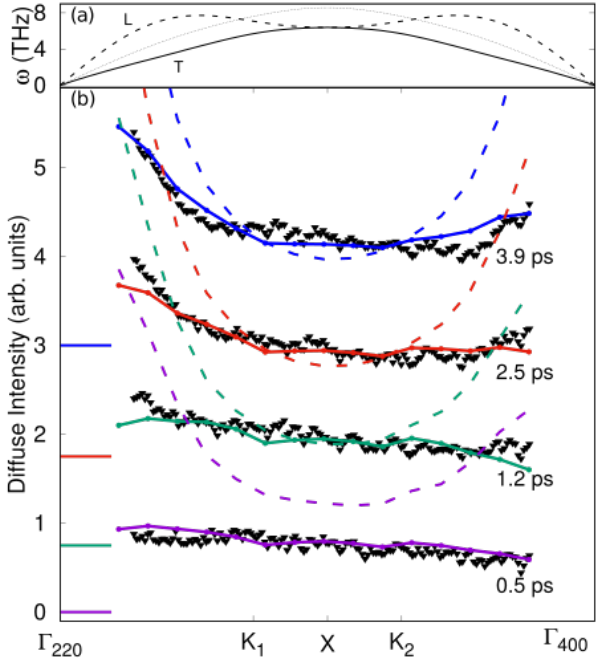

**Fig. 2.** Calculated and measured Ni phonon properties along the $\Gamma_{220}$-X-$\Gamma_{400}$ line. (a) Calculated phonon energy dispersions for one longitudinal (L, drawn line) and two transverse (T, dashed line, $T_{perp}$ dotted line) phonon branches, along the phonon linewidths due to electron-phonon and phonon-phonon interactions. Only the L and T branches are probed experimentally. (b) Measured (symbols) and theoretical (lines) diffuse scattering intensity for the indicated pump-probe time delays. Dashed lines were calculated for a ferromagnetic electronic structure while for the drawn lines it was assumed that the electronic structure becomes non-magnetic after 300 fs as described in the text.

This confirms the well-known fact that Ni demagnetizes upon ultrafast laser excitation within several 100 fs [5,6,20]. The collapsing exchange splitting on demagnetization changes the electronic structure of Ni [20,21]. This ultrafast modification of minority and majority-spin electron distributions induces an ultrafast adjustment of the spin-dependent e-ph coupling. In ferromagnetic Ni, the e-ph coupling is mainly caused by minority spin *d*-electrons close to the Fermi level while the e-ph coupling strength from majority *sp*-electrons is two orders of magnitude smaller. It is expected that the ultrafast demagnetization of Ni leads to an increase of minority-spin *sp*-electrons. These time-dependent changes of the e-ph coupling strength due to demagnetization are accounted for in a simplified manner (drawn lines in Fig. 2(b)) by assuming a ferromagnetic e-ph coupling for the first 300 fs after laser excitation [20] and a non-magnetic one afterwards, with a reduction of the e-ph coupling by a factor of four. This results in an excellent agreement with the experimental measurements for all time delays and phonon modes. It is important to note that only one single global scaling factor has been applied to match experimental and theoretical diffuse scattering intensities which demonstrates the predictive power of our model.

This allows us to gain further insights in the competition between the microscopic e-ph and ph-ph energy-transfer mechanisms and the energy flow in the laser-excited non-equilibrium state. Figure 3 summarizes the computed energy transfer rates between electrons and phonons (e-ph in Fig. 3(a)) and within the phonon system (ph-ph in Fig. 3(b)). One can clearly observe that the both e-ph and ph-ph energy transfer rates display pronounced dependences on the phonon momentum and delay time.

At early times the e-ph energy transfer rate is very high and even further enhanced due to the four-times higher e-ph coupling in the ferromagnetic state compared to the non-magnetic electronic structure after ~300 fs. The energy transfer is strongest near the BZ boundary. Phonon modes near the BZ center at all times display a positive e-ph energy transfer rate. This is reflected in a monotonous increase of the diffuse intensity with delay time at 0.25Γ-X as shown in the inset of Fig. 3 for both experiment (symbols) and theory (lines). In contrast the e-ph energy transfer at the BZ boundary changes sign leading to a broad maximum in the diffuse intensity at X near 2 ps delay time (inset of Fig. 3). At this point the direction of the energy flow is reversed and high-momentum phonons transfer energy back to electrons as depicted schematically in Fig. 1(a). This effect is enhanced by the relatively small ph-ph energy transfer rate, shown in Fig. 3(b) for three-phonon scattering processes. Note that in Fig. 3(b) energy transfer can also involve phonons away from the $\Gamma_{220}$-X-$\Gamma_{400}$ high-symmetry line. The ph-ph energy transfer rate also shows a momentum dependence. We find energy losses near the X-point and energy gains near Γ at all delay times. However, the small values of the ph-ph energy transfer rates near the X-point demonstrates that during the probed time delays the diffuse intensity near the BZ boundary indeed directly reflects the e-ph energy flow.

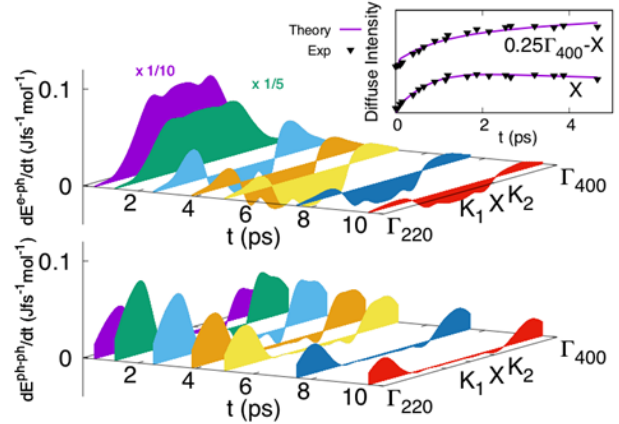

**Fig. 3.** Calculated changes of the electron-phonon (e-ph) (a) and phonon-phonon (ph-ph) (b) energy transfer rates at different time delays along the $\Gamma_{220}$-X-$\Gamma_{400}$ line for demagnetizing Ni. The inset shows the calculated and measured diffuse intensity as function of the time delay at two positions in reciprocal space.

Based on the good agreement between measured and calculated diffuse intensities we can now determine separately the phonon occupations and the energy content of the L- and T-modes. Figure 4 shows, as an example, for a delay time of 1.5 ps the (normalized) deviation from an equilibrium phonon distribution $\Delta n^n/n^{eq} = (n^{neq} - n^{eq})/n^{eq}$ with $n^{eq}$ the thermal equilibrium occupation along the Γ-K-X high-symmetry line. In Fig. 4(a) this quantity is overlaid with a color-code (blue: under-populated; red: over-populated) on the dispersion relation for the L- and T-branches. Figures 4(b,c), and (d,e) show corresponding normalized differential energy distributions $\Delta E/E^{eq} = (E^{neq} - E^{eq})/E^{eq}$ as a function of phonon momentum and frequency, respectively. These graphs clearly demonstrate that for the given delay time the energy transferred from the electron system to the lattice is predominantly stored in high-momentum phonons. Moreover, it is worth noting that the phonon momentum, not the phonon energy, determines the e-ph coupling strength and thus the phononic non-equilibrium. This is seen most clearly in the behavior of the L-branch, which exhibits a non-monotonic dispersion relation. Both, $\Delta n$ and $\Delta E$ increase monotonically with phonon momentum, but exhibit an apparently "irregular" behavior as a function of phonon energy/frequency at higher phonon energies. Here it is important to mention that although Fig. 4(a) is computed theoretically, Figs. 4(b,c), and (d,e) are extracted from the experimental diffuse intensities. However, due to the noise in the experimental data, we have used an iterative process, where starting from the theoretical L-branch component we obtain the experimental T-branch component from the data. Then by using the experimental data and the experimentally obtained T-branch component we can extract the experimental L-branch component.

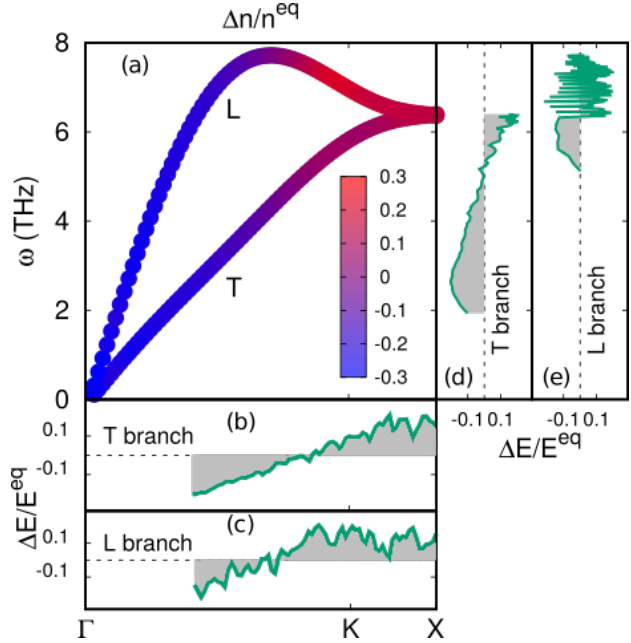

**Fig. 4**. Phonon occupation at 1.5 ps time delay. (a) Calculated wavevector, frequency and polarization dependent non-equilibrium (neq) phonon occupations relative to the equilibrium (eq) case (b, c) Phonon wavevector dependence of the experimentally determined non-equilibrium energy in the respective phonon modes relative to equilibrium. (d, e) Phonon frequency dependence of the experimentally determined non-equilibrium energy in the respective phonon modes relative to equilibrium.

It is striking that at no time do we observe phonon populations that correspond to those expected for thermal equilibrium, $n^{eq}$. While this is expected for sub-ps times, it is surprising that even after several ps it is not possible to define one global phonon temperature. We observe that the deviation from equilibrium is already significant 0.5 ps after laser excitation, showing a large energy accumulation ($n^{neq} > n^{eq}$) at the phonon "hot spots", i.e. high-wavevector phonons with strong e-ph coupling, while limited energy has flowed into the region around the BZ center ($n^{neq} < n^{eq}$). This behavior is further enhanced up to about 2 ps at which time the backflow of energy from the phonon hot spots to the electronic system sets in. This process provides, surprisingly, the most effective channel for lattice equilibration. The alternative path of lattice relaxation, i.e ph-ph energy transfer, does not contribute significantly, as it takes several 10 ps, which is much longer than the probed temporal range in this study.

*Conclusions*. We have presented a combined experimental and theoretical study of the energy dynamics during the first 5 ps following femtosecond laser excitation of a Ni film. We quantify the deviations of the phonon occupations relative to thermal equilibrium and find the results are in stark contrast to what is commonly expected from a two-temperature model analysis. Our results clearly demonstrate that during the whole time the lattice remains out-of-equilibrium even for such a simple system with only one atom per unit cell. Initially the energy delivered by the laser into the electronic systems is shared inhomogeneously with the lattice via the electron-phonon coupling with a positive energy balance from the electrons toward the lattice. This leads to phonon regions which are largely heated (BZ edges) compared with other regions (BZ center). As a consequence, an entangled energy flow between different phonon modes and the electronic system is initiated. At larger timescales the *hot* phonon modes with large energy densities start losing energy while those with smaller local energy densities keep gaining energy. We find that the energy flow from the electrons toward the lattice has an explicit dependence on the magnetic character of the system, and magnetization changes entails significant modification of the lattice dynamics. Our results demonstrate a robust and straightforward way to disentangle the complex non-equilibrium interplay between electrons and phonons that can be extended to more complex materials.

*Acknowledgments*. Theory work has been funded through the Swedish Research Council (VR), the K. and A. Wallenberg Foundation (Grant No. 2015.0060), the European Union's Horizon2020 Research and Innovation Programme under Grant Agreement No. 737709, and the Czech Science Foundation (Grant No. 15-08740Y). We also acknowledge support from the Swedish National Infrastructure for Computing (SNIC). The UED work was performed at SLAC MeV-UED, which is supported in part by the DOE BES SUF Division Accelerator & Detector R&D program, the LCLS Facility, and SLAC under contract Nos. DE-AC02-05-CH11231 and DE-AC02-76SF00515. KST and MHvH acknowledge support by the German Research Council DFG through project No. 278162697-SFB 1242 "Non-equilibrium dynamics of condensed matter in the time domain" (project BO4 and C01).

**References**
1. A. Kimel, A. Kirilyuk, P. A. Usachev, R. V. Pisarev, A. M. Balbashov, and Th. Rasing. Ultrafast non-thermal control of magnetization by instantaneous photomagnetic pulses. Nature 435, 655–657 (2005).
2. F. Schmitt, P. S. Kirchmann, U. Bovensiepen, R. G. Moore, L. Rettig, M. Krenz, J.-H. Chu, N. Ru, L. Perfetti, D. H. Lu, M. Wolf, I. R. Fisher, and Z.-X. Shen, Transient electronic structure and melting of a charge density wave in TbTe$_3$. Science 321, 1649–1652 (2008).
3. I. Radu, K. Vahaplar, C. Stamm, T. Kachel, N. Pontius, H. A. Dürr, T. A. Ostler, J. Barker, R. F. L. Evans, R. W. Chantrell, A. Tsukamoto, A. Itoh, A. Kirilyuk, Th. Rasing, and A. V. Kimel. Transient ferromagnetic-like state mediating ultrafast reversal of antiferromagnetically coupled spins. Nature 472, 205–208 (2011).
4. V. R. Morrison, R. P. Chatelain, K. L. Tiwari, A. Hendaoui, A. Bruhacs, M. Chaker, and B. J. Siwick, A Photoinduced Metal-Like Phase of Monoclinic VO$_2$ Revealed by Ultrafast Electron Diffraction. Science 346, 445 (2014).
5. E. Beaurepaire, J.-C. Merle, A. Daunois, and J.-Y. Bigot, Ultrafast spin dynamics in ferromagnetic nickel. Phys. Rev. Lett. 76, 4250–4253 (1996).
6. B. Koopmans, G. Malinowski, F. Dalla Longa, D. Steiauf, M. Fähnle, T. Roth, M. Cinchetti, and M. Aeschlimann, Explaining the paradoxical diversity of ultrafast laser-induced demagnetization. Nature Materials 9, 259–265 (2010).
7. T. Chase, M. Trigo, A. H. Reid, R. Li, T. Vecchione, X. Shen, S. Weathersby, R. Coffee, N. Hartmann, D. A. Reis, X. J. Wang, and H. A. Dürr, Ultrafast electron diffraction from non-equilibrium phonons in femtosecond laser heated Au films. Appl. Phys. Lett. 108, 041909 (2016).

8. L. Stojchevska, I. Vaskivskyi, T. Mertelj, P. Kusar, D. Svetin, S. Brazovskii, and D. Mihailovic, Ultrafast Switching to a Stable Hidden Quantum State in an Electronic Crystal. Science 344, 177–180 (2014).
9. I. Vaskivskyi, J. Gospodaric, S. Brazovskii, D. Svetin, P. Sutar, E. Goreshnik, I. A. Mihailovic, T. Mertelj, and D. Mihailovic, Controlling the metal-to-insulator relaxation of the metastable hidden quantum state in 1T-TaS$_2$. Sci. Adv. 1, e1500168 (2015).
10. M. Först, C. Manzoni, S. Kaiser, Y. Tomioka, Y. Tokura, R. Merlin, and A. Cavalleri, Nonlinear phononics as an ultrafast route to lattice control. Nat. Phys. 7, 854–856 (2011).
11. M. Mitrano A. Cantaluppi, D. Nicoletti, S. Kaiser, A. Perucchi, S. Lupi, P. Di Pietro, D. Pontiroli, M. Riccò, S. R. Clark, D. Jaksch, and A. Cavalleri. Possible light-induced superconductivity in $K_3C_{60}$ at high temperature. Nature. 530, 461–464 (2016).
12. R. Mankowsky A. Subedi, M. Först, S. O. Mariager, M. Chollet, H. T. Lemke, J. S. Robinson, J. M. Glownia, M. P. Minitti, A. Frano, M. Fechner, N. A. Spaldin, T. Loew, B. Keimer, A. Georges, and A. Cavalleri. Nonlinear lattice dynamics as a basis for enhanced superconductivity in $YBa_2Cu_3O_{6.5}$. Nature 516, 71–73 (2014).
13. S. I. Anisimov, B. L. Kapeliovich, and T. L. Perel'man, Electron Emission from Metal Surfaces Exposed to Ultrashort Laser Pulses. JETP 39, 375 (1974). [Zh. Eksp. Teor. Fiz. 66, 776 (1974)].
14. P. B. Allen, Theory of Thermal Relaxation of Electrons in Metals. Phys. Rev. Lett. 59, 1460 (1987).
15. Jhih-An Yang, S. Parham, D. Dessau, and D. Reznik, Novel Electron-Phonon Relaxation Pathway in Graphite Revealed by Time-Resolved Raman Scattering and Angle-Resolved Photoemission Spectroscopy. Sci. Rep. 7, 40876 (2016)
16. T. Henighan, M. Trigo, S. Bonetti, P. Granitzka, D. Higley, Z. Chen, M. P. Jiang, R. Kukreja, A. Gray, *et al*. Generation mechanism of terahertz coherent acoustic phonons in Fe. Phys. Rev. B 93, 220301(R) (2016).
17. L. Waldecker, R. Bertoni, R. Ernstorfer, and J. Vorberger, Electron-Phonon Coupling and Energy Flow in a Simple Metal beyond the Two-Temperature Approximation. Phys. Rev. X 6, 021003 (2016).
18. L. Waldecker, R. Bertoni, H. Hübener, T. Brumme, T. Vasileiadis, D. Zahn, A. Rubio, and R. Ernstorfer, Momentum-Resolved View of Electron-Phonon Coupling in Multilayer $WSe_2$. Phys. Rev. Lett. 119, 036803 (2017).
19. T. Konstantinova, J. D. Rameau, A. H. Reid, O. Abdurazakov, L. Wu, R. Li, X. Shen, G. Gu, Y. Huang, L. Rettig, I. Avigo, M. Ligges, J. K. Freericks, A. F. Kemper, H. A. Dürr, U. Bovensiepen, P. D. Johnson, X. Wang, and Y. Zhu, Nonequilibrium electron and lattice dynamics of strongly correlated $Bi_2Sr_2CaCu_2O_{8+d}$ single crystals. Sci. Adv. 4, eaap7427 (2018).
20. H.-S. Rhie, H. A. Dürr, and W. Eberhardt, Phys. Rev. Lett. 90, 247201 (2003).
21. W. You, P. Tengdin, C. Chen, X. Shi, D. Zusin, Y. Zhang, C. Gentry, A. Blonsky, M. Keller, P. M. Oppeneer, H. Kapteyn, Z. Tao, and M. Murnane, Phys. Rev. Lett. 121, 077204 (2018).
22. P. Maldonado, K. Carva, M. Flammer, and P.M. Oppeneer, Theory of out-of-equilibrium ultrafast relaxation dynamics in metals. Phys. Rev. B 96, 174439 (2017).
23. K. Sokolowski-Tinten, R. Li, A. H. Reid, S. P. Weathersby, F. Quirin, T. Chase, R. Coffee, J. Corbett, A. Fry, N. Hartmann, C. Hast, R. Hettel, M. Horn von Hoegen, D. Janoschka, J. R. Lewandowski, M. Ligges, F. Meyer zu Heringdorf, X. Shen, T. Vecchione, C. Witt, J. Wu, H. A. Dürr, and X. Wang, Thickness-dependent electron-lattice equilibration in laser-excited thin Bismuth films, New J. Phys. 17, 113047 (2015).
24. K. Sokolowski-Tinten, X. Shen, Q. Zheng, T. Chase, R. Coffee, M. Jerman, R. K. Li, M. Ligges, I. Makasyuk, M. Mo, A. H. Reid, B. Rethfeld, T. Vecchione, S. P. Weathersby, H. A. Dürr, and X. J. Wang, Electron-lattice energy relaxation in laser-excited thin-film Au-insulator heterostructures studied by ultrafast MeV electron diffraction, Struct. Dyn. 4, 054501 (2017).
25. Q. Zheng, X. Shen, K. Sokolowski-Tinten, R. K. Li, Z. Chen, M. Z. Mo, Z. L. Wang, S. P. Weathersby, J. Yang, M. W. Chen, and X. J. Wang, Dynamics of electron-phonon coupling in bicontinuous nanoporous gold, J. Phys. Chem. C, 122, 16368−16373 (2018).
26. R. P. Chatelain, V. R. Morrison, B. L.M. Klarenaar, and B. J. Siwick, Coherent and Incoherent Electron-Phonon Coupling in Graphite Observed with Radio-Frequency Compressed Ultrafast Electron Diffraction. Phys. Rev. Lett. 113, 235502 (2014).
27. M. Trigo, M. Fuchs, J. Chen, M. P. Jiang, M. Cammarata, S. Fahy, D. M. Fritz, K. Gaffney, S. Ghimire, A. Higginbotham, S. L. Johnson, M. E. Kozina, J. Larsson, H. Lemke, A. M. Lindenberg, G. Ndabashimiye, F. Quirin, K. Sokolowski-Tinten, C. Uher, G. Wang, J. S. Wark, D. Zhu, and D. A. Reis, Fourier-transform inelastic X-ray scattering from time- and momentum-dependent phonon–phonon correlations. Nat. Phys. 9, 790 (2013).
28. M. Harb, H. Enquist, A. Jurgilaitis, F. T. Tuyakova, A. N. Obraztsov, and J. Larsson, Phonon-phonon interactions in photoexcited graphite studied by ultrafast electron diffraction. Phys. Rev. B 93, 104104 (2016).
29. M. J. Stern, L. P. René de Cotret, M. R. Otto, R. P. Chatelain, J.-P. Boisvert, M. Sutton, and B. J. Siwick, Mapping Momentum-Dependent Electron-Phonon Coupling and Nonequilibrium Phonon Dynamics with Ultrafast Electron Diffuse Scattering. Phys. Rev. B 97, 165416 (2018).
30. S. P. Weathersby, G. Brown, M. Centurion, T. F. Chase, R. Coffee, J. Corbett, J. P. Eichner, J. C. Frisch, A. R. Fry, M. Gühr, N. Hartmann, C. Hast, R. Hettel, R. K. Jobe, E. N. Jongewaard, J. R. Lewandowski, R. K. Li, A. M. Lindenberg, I. Makasyuk, J. E. May, D. McCormick, M. N. Nguyen, A. H. Reid, X. Shen, K. Sokolowski-Tinten, T. Vecchione, S. L. Vetter, J. Wu, J. Yang, H. A. Dürr, and X. J. Wang, Mega-electron-volt ultrafast electron diffraction at SLAC National Accelerator Laboratory. Rev. Sci. Instrum. 86, 073702 (2015).
31. M. Holt, Z. Wu, H. Hong, P. Zschack, P. Jemian, J. Tischler, H. Chen, and T.-C. Chiang, Determination of phonon dispersions from x-ray transmission scattering: The example of silicon. Phys. Rev. Lett. 83, 3317 (1999).
32. Ruqing Xu and Tai C. Chiang, Determination of phonon dispersion relations by X-ray thermal diffuse scattering. Z. Kristallogr. 220, 1009–1016 (2005).
33. M. Z. Mo, Z. Chen, R. K. Li, M. Dunning, B. B. L. Witte, J. K. Baldwin, L. B. Fletcher, J. B. Kim, A. Ng, R. Redmer, A. H. Reid, P. Shekhar, X. Z. Shen, M. Shen, K. Sokolowski-Tinten, Y. Y. Tsui, Y. Q. Wang, Q. Zheng, X. J. Wang, S. H. Glenzer, Heterogeneous to homogeneous melting transition visualized with ultrafast electron diffraction. Science 360, 6396 (2018).